\documentclass[prd,twocolumn,showpacs,floatfix,amsmath,nofootinbib,amssymb,floatfix]{revtex4}
\usepackage{graphicx,color,dcolumn,booktabs,bm,multirow}
\usepackage{longtable,lscape}
\usepackage{txfonts}
\usepackage{overpic}
\usepackage{amssymb}
\usepackage{indentfirst}
\usepackage{feynmf}
\usepackage{slashed}
\usepackage{cases}
\usepackage{color}
\usepackage{multirow}
\usepackage{graphicx,color,dcolumn,booktabs,bm}
\usepackage{ulem}

\usepackage[colorlinks,
            citecolor=blue,
            anchorcolor=red,
            linkcolor=blue,
            filecolor=red,
            runcolor=red,
            urlcolor=blue,
            frenchlinks=red]{hyperref}

\newcommand{\eq}{\begin{eqnarray}}
\newcommand{\en}{\end{eqnarray}}
\begin{document}

\title{Towards the decay properties of deuteron-like state $d_{N\Omega}$}
\author{Cheng-Jian Xiao$^{1,2}$\footnote{Corresponding author}}\email{xiaocj@ihep.ac.cn}
\author{Yu-Bing Dong$^{1,2,3}$}\email{dongyb@ihep.ac.cn}
\author{Thomas Gutsche$^{4}$}\email{thomas.gutsche@uni-tuebingen.de}
\author{Valery E. Lyubovitskij$^{4,5,7}$}\email{valeri.lyubovitskij@uni-tuebingen.de}
\author{Dian-Yong Chen$^{6}$}\email{chendy@seu.edu.cn}

\affiliation{$^1$Institute of High Energy Physics, Chinese
     Academy of Sciences, Beijing 100049, People's Republic of China\\
$^2$ Theoretical Physics Center for Science Facilities (TPCSF), CAS, Beijing 100049, China\\
$^3$ School of Physical Sciences, University of Chinese
Academy of Sciences, Beijing 101408, China\\
$^4$ Institut f\"ur Theoretische Physik, Universit\"at
     T\"ubingen, Kepler Center for Astro and Particle Physics,
     Auf der Morgenstelle 14, D-72076 T\"ubingen, Germany\\
$^5$Departamento de F\'\i sica y Centro Cient\'\i fico
    Tecnol\'ogico de Valpara\'\i so-CCTVal, Universidad
    T\'ecnica Federico Santa Mar\'\i a, Casilla 110-V, Valpara\'\i so, Chile\\
$^6$ School of Physics, Southeast University,
    Nanjing 210094, People's Republic of China\\
$^7$ Department of Physics, Tomsk State University, Tomsk 634050, Russia
    }

\begin{abstract}

Recent lattice QCD calculations showed that a $d_{N\Omega}$ similar to the deuteron with baryon number $B=2$ and with a small binding
energy might exist. In this work we propose a hadronic molecular approach to study the dynamical properties of this exotic state.
We employed a phenomenological Lagrangian approach to describe the coupling of the $d_{N\Omega}$ to its constituents and the
strong decays into conventional hadrons, $d_{N\Omega} \to \Lambda \Xi$ and $d_{N\Omega} \to \Sigma \Xi$. Predictions
for the sum of the decay rates are in the range of a few hundred keV. In addition, we find that the $d_{N\Omega} \to \Lambda \Xi$ mode
is dominant, preferably searched for in a future Relativistic Heavy Ion Collider (RHIC) experiment.

\end{abstract}

\date{\today}

\maketitle

\section{Introduction}

Since the discovery of the $X(3872)$ state in 2003~\cite{Choi:2003ue} the study of exotic resonances with heavy flavors, like $X(3872)$,
$Z_c(3900)$ or the $P_c$ states, turns out to be extremely important in unravelling their unusual internal structure both in
theoretical and experimental investigations~\cite{Patrignani:2016xqp}. In particular, many experimental efforts at world-wide
facilities (like BEPC II, BELLE, CERN, JLab, LHCb, etc.) have been carried out for hunting and identifying those
exotics~\cite{Patrignani:2016xqp,Ablikim:2013mio,Ablikim:2013wzq,Aaij:2015tga,Aaij:2019vzc,Ali:2019lzf}.
Numerous theoretical calculations were also devoted to the understanding of those unusual hadron states with respect to their composite
structure, mass spectrum and decay properties [for a detailed list of references and reviews, see, e.g.,
Refs.~\cite{Patrignani:2016xqp,Maiani:2004vq,Swanson:2006st,Lebed:2016hpi,Dong:2017gaw,Guo:2017jvc,Chen:2016qju,Esposito:2016noz,
Karliner:2017qhf,Liu:2019zoy,Guo:2019twa}]. Different interpretations have been proposed and developed in the literatures: hadronic
molecular scenarios, multiquark states -- tetraquark or pentaquark configurations, kinematic triangle singularities, and scattering cusps,
among some others.

Multiquark states can be realized not only as four-quark (meson sector) and five-quark (baryon sector) systems,
but also as six-quark state. Correspondingly, there are meson-meson, meson-baryon and baryon-baryon molecular states. The deuteron,
discovered in 1931, is the prototype of a baryon-baryon molecular state, mainly residing in a proton-neutron configuration with a
weak binding energy of $E_{b}\simeq~2.22~\rm{MeV}$. Dyson and Xuong  were the first to study nonstrange two-baryon systems
in terms of $SU(6)$ even before the quark model was established~\cite{Dyson:1964xwa}. The H-particle, originally proposed by
Jaffe~\cite{Jaffe:1976yi}, and other candidates like  the $d^*$, were searched for in experiments for a very long time. Recently, the
nonstrange resonance $d^*(2380)$ was observed and confirmed by the WASA@COSY
collaboration~\cite{CELSIUS-WASA,Adlarson:2011bh,Adlarson:2012fe,Adlarson:2014pxj}. So far, the understanding of the nature of
the $d^*(2380)$ resonance is not conclusive. The three-diquark state\cite{shi:2019iu}, compact six-quark
state~\cite{Yuan:1999pg,Huang:2014kja,Dong:2015cxa, Dong:2016rva,Dong:2017geu} or hadronic molecule structure~\cite{Gal:2014zia}
are three possible interpretations (see the review article~\cite{Clement:2016vnl}).

Possible nucleon-hyperon states, with baryon number $B=2$, have also been studied in the
literature~\cite{Jaffe:1976yi,Goldman:1989zj,Zhang:2000sv,Pang:2003ty,Liu:2011xc,Froemel:2004ea}.
The $N\Omega$ state is a typical example among them as it is believed to be bound. The first investigation of a six-quark system
with strangeness $S=-3$ was done by Goldman~{\it et al.} using the relativistic quark model~\cite{Goldman:1987ma}.
They proposed a bound $S$-wave $N\Omega$  state with total angular momentum $J=1$ or 2. Later on, in Ref.~\cite{Oka:1988yq} it was
pointed out that the treatment with a single $N\Omega$ channel cannot lead to a bound state, since there is no quark exchange effect
in this channel. When considering the coupled channels, like $N\Omega-\Lambda\Xi^\ast-$
$\Sigma\Xi^\ast-\Sigma^\ast\Xi-\Sigma^\ast\Xi^\ast$, a bound state might exist. The $N\Omega$ system was also studied in a quark
delocalization and color screening model (QDCSM), where the bound state can be obtained both for the single $N\Omega$
or coupled channel configurations~\cite{Pang:2004mm}. The predicted masses are $M=2566~\rm{MeV}$ and $M=2549~\rm{MeV}$ for the
two cases, respectively. A further analysis of the QDCSM was recently performed in Ref.~\cite{Huang:2015yza}, and the updated results
were consistent with the previous ones. A bound $N\Omega$ state is also supported by chiral quark model calculations,
where the binding energy varies from ten to around one hundred MeV  depending on the specific approach~\cite{Li:2000cb,Huang:2015yza}.
Moreover, Ref.~\cite{Sekihara:2018tsb} found  a quasibound $N\Omega$ state with a pole at $E_{pole}=2611.3-0.7i~\rm{MeV}$ based on a
meson exchange model.

Besides those model calculations, recently, a lattice calculation for the $d_{N\Omega}$  system was performed by the HAL QCD
Collaboration~\cite{Inoue:2011ai,Iritani:2018sra}.  As a result they reported that an  $S$-wave $d_{N\Omega}$ with $J^{P}=2^+$
and with deuteronlike binding energy of $E_b=2.46~\rm{MeV}$ indeed does exit.  The HAL QCD Collaboration performed their
lattice simulations for nearly physical quark masses corresponding to pseudoscalar masses of $m_{\pi}\simeq 146~\rm{MeV}$ and
$m_{K}\simeq~525~\rm{MeV}$.  The possible strong short range attraction in the proton-$\Omega$ system can also be accessed by the
momentum correlation of $p\Omega$ emission in relativistic heavy ion collisions~\cite{Morita:2016auo}. The corresponding measurement
has been carried out by the STAR Collaboration at Relativistic Heavy Ion Collider (RHIC) using the Au+Au collision~\cite{STAR:2018uho}. The results slightly favor a
bound $d_{N\Omega}$  with a binding energy of about $27~\rm{MeV}$. Besides the first work in Ref.~\cite{Morita:2016auo} the authors
extended their analysis on the pair momentum correlation functions in \cite{Morita:2019rph}.

To check for the existence of a $d_{N\Omega}$, a direct search for a signal in the invariant mass of the final decay channels is
necessary. Therefore, also a theoretical calculation on the decay properties of the $d_{N\Omega}$ is needed.  In this work we consider
the $d_{N\Omega}$ as a loosely bound state of a nucleon and a $\Omega$ with a value for the binding energy set by the lattice
calculation. Then we employ an effective Lagrangian approach to calculate the strong decays. It should be mentioned that the
phenomenological Lagrangian approach is a reasonable method to describe the properties of weakly bound states. We have
successfully applied it to a wide range of exotic resonances, like $D_{s0}^\ast(2317)$, $X(3872)$, $Z_c(3900)$, and $Y(4260)$
in the meson sectors~\cite{Faessler:2007gv,Faessler:2007us,Dong:2008gb,Dong:2009yp,Dong:2009uf,Branz:2009yt,
Dong:2013iqa,Chen:2015igx,Dong:2013kta,Xiao:2016mho,Xiao:2016hoa}, and for $\Lambda_c(2940)$, $\Sigma_c(2800)$, $\Omega(2012)$,
and $P_c$  in the baryon sector~\cite{Dong:2010xv,Dong:2009tg,Lu:2016nnt,Gutsche:2019mkg, Xiao:2019mst,Gutsche:2019imd}. We also
employed this method to study deuteron properties~\cite{Dong:2008mt,Liang:2013pqa}. Since the binding of this $d_{N\Omega}$ state is
expected to be similar to that of the deuteron, we expect that our phenomenological Lagrangian approach will result in
reasonable predictions for the strong decay properties of the $d_{N\Omega}$.

This paper is organized as follows. In Sec.~\ref{section:2}, we discuss the setup of the hadronic structure of the $d_{N\Omega}$ bound
state and follow up with the formalism of the strong decay modes in the context of an effective Lagrangian approach.
Section~\ref{section:3} is devoted to the numerical evaluation and discussion of the strong decays of this $N\Omega$ molecular state.
Finally, a short summary will be given in Sec.~\ref{section:4}.

\section{Strong decays of the $d_{N\Omega}$  \label{section:2}}

In the following we assume that the $d_{N\Omega}$ is a loosely bound state of a nucleon and an $\Omega^-$ hyperon.
Following the results of the recent lattice calculations~\cite{Iritani:2018sra}, the quantum numbers of
the $d_{N\Omega}$ weakly bound state are chosen as $I(J^P)=\frac12(2^+)$. The bound state has two isospin components,
$p\Omega$ for  $I_3=1/2$ and $n\Omega$ for $I_3=-1/2$. To set up a framework for the treatment of a bound state of
two hadrons we construct a phenomenological Lagrangian describing the interaction of the $d_{N\Omega}$ with its
constituents as
\eq
\mathcal L&=&{g_{_{\scriptstyle d_{N\Omega}}}}  d_{N\Omega}^{\mu\nu\dagger}
    \int dy \, \Phi(y^2) \, \bar\Omega_\mu^c(x+\omega_{_{\scriptstyle N\Omega}}y)
    \gamma_\nu N(x-\omega_{_{\scriptstyle \Omega N}} y)\nonumber\\
&+&\text{h.c.}\, ,
\label{eq:lag-d-nomega}
\en
where $\psi^c=C\bar \psi^T$, $\bar\psi^c=\psi^TC$, and $\bar\psi^c_1 \gamma^\mu \psi_2 = \bar\psi^c_2 \gamma^\mu \psi_1$.
Here $C=i\gamma^2\gamma^0$ is the charge-conjugation matrix, superscript $T$ denotes the transposition and $\omega_{ij}= m_i/(m_i+m_j)$
is the hadron mass fraction parameter, where $m_i$ is the mass of the $i$th particle. To describe the distribution of the constituents
in the hadronic molecular system, we introduce the correlation function $\Phi(y^2)$, which, in addition, plays the role to render
the Feynman diagrams ultraviolet finite. Note that $\Phi(y^2)$ is related to its Fourier transform in momentum space $\tilde\Phi(-p^2)$ as:
\eq\label{eq:corre-func}
\Phi(y^2)=\int \frac{d^4 p}{(2\pi)^4}e^{-ipy}\tilde\Phi(-p^2)\,,
\en
where $p = \omega_{N\Omega} p_{\Omega} -\omega_{\Omega N} p_N$ is the Jacobi momentum. Here $\tilde\Phi(-p^2)$ is the correlation
function describing the distribution of constituents in the molecular state. It was widely and successfully used in the investigation
of hadronic molecules~\cite{Chen:2015igx,Faessler:2007gv,Faessler:2007us,Branz:2009yt,Xiao:2016mho,Xiao:2016hoa,Dong:2008mt}.
For simplicity $\tilde\Phi$ is chosen as a Gaussian-like form $\tilde\Phi(-p^2) = \exp(p^2/\Lambda^2)$, where $\Lambda$ is the
model parameter, which has dimension of mass and defines a scale for the distribution of the constituents inside the molecule.
All calculations are performed in Euclidean space after Wick transformation for loop and all external momenta:
$p^\mu=(p^0,\vec{p\,}) \to p^\mu_E=(p^4,\vec{p\,})$ with $p^4 = -ip^0$. In Euclidean space the Gaussian correlation function provides
that all loop integrals are ultraviolet finite.

\begin{figure}[hbt!]
\includegraphics[scale=0.5]{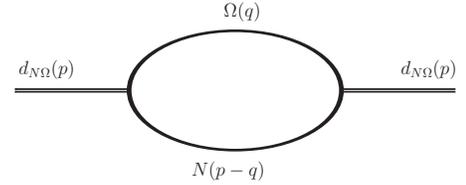}
\caption{Mass operator of the $d_{N\Omega}$. \label{fig:mass-operator}}
\end{figure}

The coupling constant $g_{d_{N\Omega}}$ in Eq.~(\ref{eq:lag-d-nomega}) is determined using the Weinberg-Salam compositeness
condition~\cite{Salam:1962ap,Weinberg:1962hj,Hayashi:1967hk,Efimov:1993ei}. This condition means that the probability to find
the dressed bound state as a bare (structureless) state is equal to zero. It also means that the corresponding wave function
renormalization constant $Z$  is set to be zero. In the case of the $d_{N\Omega}$ the compositeness condition reads:
\eq
Z_{d_{N\Omega} } = 1
- \frac{\partial\Sigma^{(1)}_{d_{N\Omega}}(m^2_{d_{N\Omega}})}
{\partial m^2_{d_{N\Omega}}}=0\,,
\label{eq:compositeness-con-1}
\en
where $\Sigma^{(1)}_{d_{N\Omega}}(m^2_{d_{N\Omega}})$ is the nonvanishing part of the mass operator of the $d_{N\Omega}$  having
spin-parity $2^+$.  In Fig.~\ref{fig:mass-operator} we display the diagram contributing to the mass operator of the $d_{N\Omega}$.
Note that the respective mass operator of the $2^+$ hadron is given by the rank-4 tensor $\Sigma_{\mu\nu\alpha\beta}$ sandwiched
by the polarization vectors $\epsilon^{(\lambda)}_{\mu\nu}(p)$ for the spin $2^+$ tensor:
\eq
\hat{\Sigma}(p) =
\epsilon^{\dagger (\lambda)}_{\mu\nu}(p) \, \Sigma^{\mu\nu\alpha\beta}(p) \,
\epsilon^{(\lambda)}_{\alpha\beta}(p) \,.
\en
The polarization vector $\epsilon^{(\lambda)}_{\mu\nu}(p)$ obeys the conditions of symmetry $\epsilon^{(\lambda)}_{\mu\nu}(p) =
\epsilon^{(\lambda)}_{\nu\mu}(p)$, transversality $p^\mu \, \epsilon^{(\lambda)}_{\mu\nu}(p) = 0$,
and tracelessness $g^{\mu\nu} \, \epsilon^{(\lambda)}_{\mu\nu}(p) = 0$.

The expression for the mass operator $\Sigma_{\mu\nu\alpha\beta}$ reads as follows:
\eq
\Sigma_{\mu\nu\alpha\beta}&=&
g_{d_{N\Omega}}^2\int\frac{d^4 q}{(2\pi)^4i}
\tilde\Phi^2(-(q-w_{\Omega N}p)^2) \nonumber \\
&\times&
{\rm Tr}\bigg[ \gamma_\nu \, S_{\mu\alpha}(q,m_\Omega) \, \gamma_\beta \,
S(q-p,m_N)  \bigg]
\,, \label{eq:compositeness-con-3}
\en
where $S$ and $S_{\mu\alpha}$ are the free fermion propagators for spin-$\frac{1}{2}$ and spin-$\frac{3}{2}$ particles with
\eq
S(p,m) &=& (p\!\!\!\slash-m)^{-1} \,, \\
S_{\mu\nu}(p,m) &=& (p\!\!\!\slash-m)^{-1}  \nonumber\\
&\times& \Big(-g_{\mu\nu}+\frac{\gamma_\mu\gamma_\nu}{3}
+\frac{2p_\mu p_\nu}{3m^2}
+\frac{\gamma_\mu p_\nu-\gamma_\nu p_\mu}{3m}\Big ) \,.
\en
Using properties of the polarization vector $\epsilon^{(\lambda)}_{\mu\nu}(p)$ mentioned above $\Sigma_{\mu\nu\alpha\beta}$ can be
decomposed into the Lorentz structures $L^{(i)}_{\mu\nu\alpha\beta}$ ($i=1,\ldots,5$) multiplied by the scalar functions
$\Sigma^{(i)}(p^2)$ with:
\eq\label{mass_dNO}
\Sigma_{\mu\nu\alpha\beta}(p) &=& \sum\limits_{i=1}^5 \,
L^{(i)}_{\mu\nu\alpha\beta} \, \Sigma^{(i)}(p^2)\,,
\en
where
\eq
L^{(1)}_{\mu\nu\alpha\beta} &=&
\frac{1}{2} \biggl[ g_{\mu\alpha} g_{\nu\beta}
+ g_{\nu\alpha} g_{\mu\beta} \biggr]\,, \nonumber\\
L^{(2)}_{\mu\nu\alpha\beta} &=&
g_{\mu\nu} g_{\alpha\beta} \,, \nonumber\\
L^{(3)}_{\mu\nu\alpha\beta} &=&
\frac{1}{2} \biggl[ g_{\mu\nu} p_\alpha p_\beta
+ g_{\alpha\beta} p_\mu p_\nu \biggr] \,, \nonumber\\
L^{(4)}_{\mu\nu\alpha\beta} &=&
\frac{1}{4} \biggl[
  g_{\mu\alpha} p_\nu p_\beta
+ g_{\mu\beta}  p_\nu p_\alpha
+ g_{\nu\alpha} p_\mu p_\beta
+ g_{\nu\beta}  p_\mu p_\alpha \biggr] \,, \nonumber\\
L^{(5)}_{\mu\nu\alpha\beta} &=& p_\mu p_\nu p_\alpha p_\beta \,.
\en
As already mentioned, due to the properties of the polarization vector $\epsilon^{(\lambda)}_{\mu\nu}(p)$
only the first term in the sum of Eq.~(\ref{mass_dNO}) contributes, while the others vanish. The scalar function $\Sigma^{(1)}(p^2)$
contributing to the compositeness condition Eq.~(\ref{eq:compositeness-con-1}) is obtained from the full mass operator
$\Sigma_{\mu\nu\alpha\beta}(p)$ when acting with the following Lorentz projector
\eq
T^{\mu\nu\alpha\beta}_\perp = \frac{1}{10} \ \biggl( P^{\mu\alpha}_\perp P^{\nu\beta}_\perp
+ P^{\mu\beta}_\perp P^{\nu\alpha}_\perp \biggr) \,-\, \frac{1}{15} \ P^{\mu\nu}_\perp P^{\alpha\beta}_\perp \, .
\en
The projector $P^{\mu\nu}_\perp$ is defined as $P^{\mu\nu}_\perp = g^{\mu\nu} - p^\mu p^\nu/p^2$
and satisfies the conditions:
\eq
g_\mu^\alpha \, P^{\mu\nu}_\perp = P^{\alpha\nu}_\perp\,, \quad
g_{\mu\nu} \, P^{\mu\nu}_\perp = 3\,, \quad
p_\mu      \, P^{\mu\nu}_\perp = p_\nu \, P^{\mu\nu}_\perp = 0 \,.
\en
The full projector $T^{\mu\nu\alpha\beta}_\perp$ satisfies the following conditions
\eq
\hspace*{-.5cm}
& &p_i \, T^{\mu\nu\alpha\beta}_\perp = 0\,, \quad i = \mu,\nu,\alpha,\beta\,,
\nonumber\\
\hspace*{-.5cm}
& &L^{(1)}_{\mu\nu\alpha\beta} \, T^{\mu\nu\alpha\beta}_\perp = 1\,, \quad
   L^{(j)}_{\mu\nu\alpha\beta} \, T^{\mu\nu\alpha\beta}_\perp = 0\,, \ j=2,3,4,5\,.
\en
Finally, the required scalar function  $\Sigma^{(1)}(p^2)$ is fixed using the identity
\eq
\Sigma^{(1)}(p^2) = T^{\mu\nu\alpha\beta}_\perp \
\Sigma_{\mu\nu\alpha\beta}(p) \,.
\en

\begin{figure}[hbt!]
\begin{tabular}{cc}
\includegraphics[scale=0.45]{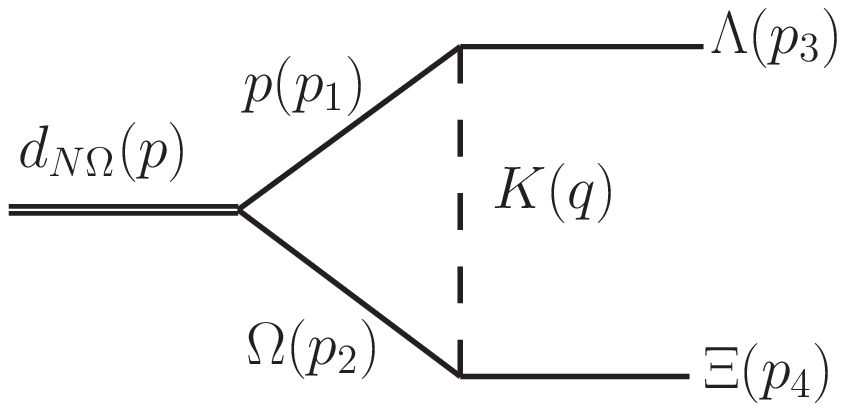}
&\hspace{-0.2cm}\includegraphics[scale=0.45]{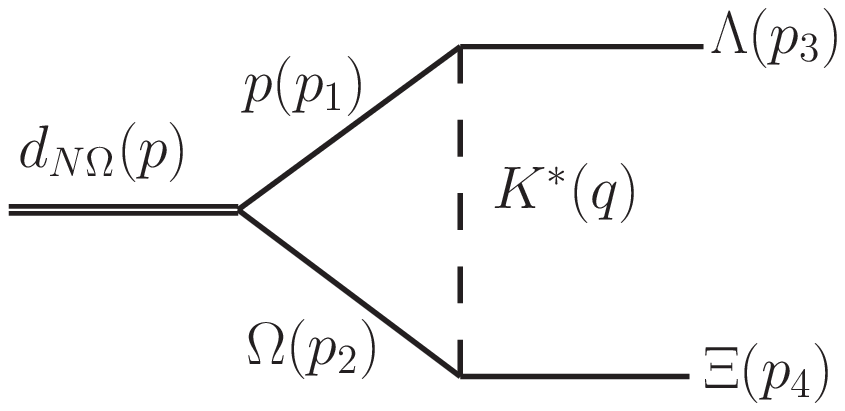}\\
(1) &(2)  \vspace{0.4cm} \\

\includegraphics[scale=0.45]{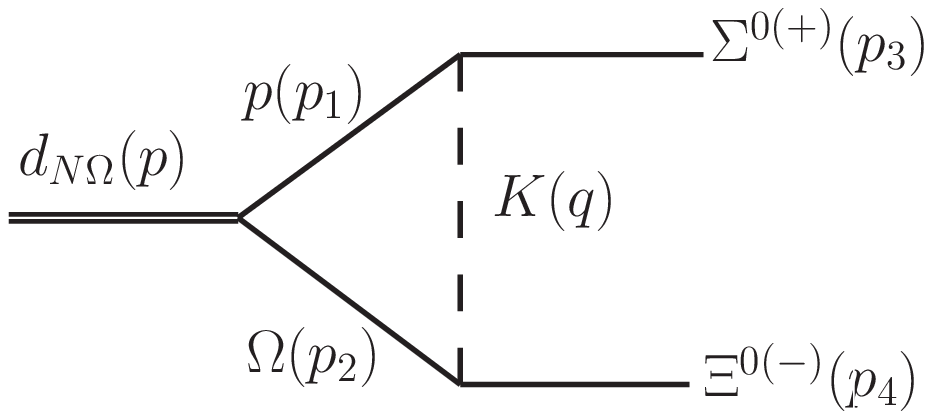}
&\hspace{-0.2cm}\includegraphics[scale=0.45]{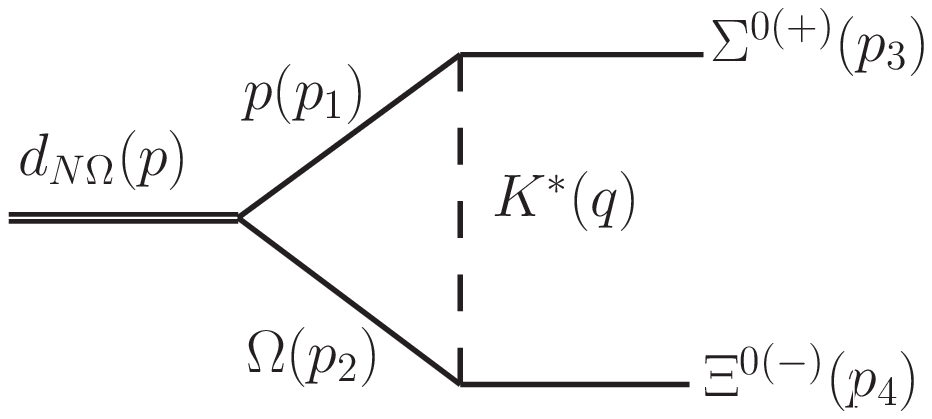}\\
(3) & (4)
  \end{tabular}
  \vspace{0.4cm}
\caption{Typical diagrams contributing to the processes $d_{N\Omega}\to \Xi\Lambda$ [diagrams (1)-(2)]
and $\Xi \Sigma$ [diagrams (3)-(4)], respectively. \label{fig:dia-xi-sig/lam} \label{fig:tri-decay}}
\end{figure}

Based on the quantum number assignment $I(J^P)=\frac12(2^+)$ of the $d_{N\Omega}$ we consider the strong decays into the baryon
pairs $\Lambda\Xi$ and $\Sigma\Xi$. In the hadronic molecular picture the decays $d_{N\Omega} \to \Lambda\Xi$ and
$d_{N\Omega} \to \Sigma\Xi$ are described by the triangle diagrams induced by the exchange of $K$ and $K^*$ mesons in the
$t$-channel. The corresponding diagrams are shown in Fig.~\ref{fig:dia-xi-sig/lam}.

To determine the matrix elements corresponding to the diagrams in Fig.~\ref{fig:dia-xi-sig/lam} we apply a phenomenological Lagrangian
including the coupling of the $d_{N\Omega}$ to its  constituents [which has been already specified in Eq.~(\ref{eq:lag-d-nomega})]
and the couplings of the constituents to the final hadrons. Thus, we need additional phenomenological Lagrangians describing the
couplings between baryons $B$ (octet and decuplet states) and mesons (pseudoscalar $P$ and vector $V$ states). In the present
calculation we use the $BBP$ and $BBV$ type Lagrangians with~\cite{Schutz:1994ue,Liu:2001ce,Machleidt:1987hj,%
Matsuyama:2006rp,Ronchen:2012eg,He:2017aps}:
\eq
\hspace*{-.5cm}
\mathcal{L}_{\Lambda NK}&=&\frac{f_{\Lambda NK}}{m_\pi}
    \bar N \gamma^\mu\gamma^5 \Lambda \partial_\mu K
\,+\, {\rm H.c.} \,, \nonumber\\
\vspace*{-.5cm}
\mathcal{L}_{\Sigma NK}&=&\frac{f_{\Sigma NK}}{m_\pi}
    \bar N \gamma^\mu\gamma^5
\hat{\Sigma} \partial_\mu K
\,+\, {\rm H.c.} \,, \nonumber\\
\hspace*{-.5cm}
\mathcal{L}_{\Lambda NK^\ast}&=&-g_{\Lambda NK^\ast}\bar N
    \big(\gamma^\mu \Lambda -
\frac{\kappa_{\Lambda NK^\ast}}{2m_N}\sigma^{\mu\nu}\Lambda \partial_\nu
    \big)K^\ast_\mu \,+\, {\rm H.c.}\,, \nonumber\\
\hspace*{-.5cm}
\mathcal{L}_{\Sigma NK^\ast}&=&-g_{\Sigma NK^\ast}\bar N
    \big(\gamma^\mu\hat{\Sigma} - \frac{\kappa_{\Sigma
    NK^\ast}}{2m_N}\sigma^{\mu\nu}\hat{\Sigma} \partial_\nu
    \big)K^\ast_\mu \,+\, {\rm H.c.}\,, \nonumber\\
\hspace*{-.5cm}
\mathcal{L}_{\Omega\Xi K}&=&\frac{f_{\Omega\Xi K}}{m_\pi}\partial_\mu K
    \bar \Omega^\mu\Xi  \,+\, {\rm H.c.}\,, \nonumber\\
\hspace*{-.5cm}
\mathcal{L}_{\Omega\Xi K^\ast}&=&
\frac{g_{\Omega\Xi K^\ast}}{m_{\rho}} \,
(\partial_\mu
K^\ast_\nu-\partial_\nu
    K^\ast_\mu){\bar\Omega^\mu}i\gamma^\nu\gamma^5\Xi
\,+\, {\rm H.c.}\,, \label{eq:lag-fin}
\en
where $\hat{\Sigma} = \overrightarrow{\Sigma}\cdot\vec\tau$.
The couplings of the octet baryons to
pseudoscalar/vector mesons are constrained by $SU(3)$-flavor symmetry
relations~\cite{Liu:2001ce,Ronchen:2012eg}:
\eq
&&f_{\Lambda NK}=-\frac{1}{\sqrt3}g_{NN\pi}(1+2\alpha_{BBP})\,,
\label{eq:coup-disc-1}\label{eq:coup-relation-begin}\\
&& f_{\Sigma NK}=g_{NN\pi}(1-2\alpha_{BBP}) \,,\label{eq:coup-disc-2} \\
&&g_{\Lambda NK^{\ast}}=-\frac{1}{\sqrt3}g_{NN\rho}(1+2\alpha_{BBV}) \,,\\
&&g_{\Sigma NK^{\ast}}=g_{NN\rho}(1-2\alpha_{BBV})\,.
\end{eqnarray}
The remaining parameter $\kappa$ in the $BBV$ coupling is fixed using the relation between vector and tensor couplings
$f_{YNK^\ast}=g_{YNK^\ast}\kappa_{YNK^\ast}$, and the relation of the tensor couplings $f_{YNK^\ast}$  to the
$f_{NN\omega}$ and $f_{NN\rho}$ couplings~~\cite{Ronchen:2012eg}:
\eq
&&f_{\Lambda NK^{\ast}}=-\frac{1}{2\sqrt3}f_{NN\omega}
    -\frac{\sqrt3}{2}f_{NN\rho},\\
&&f_{\Sigma NK^{\ast}}=-\frac12f_{NN\omega}+\frac12f_{NN\rho}.
\end{eqnarray}
For the couplings between the baryon decuplet and the pseudoscalar/vector meson octets $g_{\Delta N\pi}$ and $g_{\Delta N\rho}$, we use
$SU(3)$ symmetry constraints~\cite{Ronchen:2012eg}:
\eq
g_{\Omega\Xi K}=g_{\Delta N\pi},\quad g_{\Omega\Xi K^{\ast}}=g_{\Delta N\rho}
\label{eq:coup-relation-end}\,.
\en

\begin{table}[hbt!]
\caption{Meson-baryon coupling constants. \label{tab:value-coup}}
\begin{tabular*}{\hsize}{@{}@{\extracolsep{\fill}}ccc@{}}
 \hline
Coupling &Set I     &Set II   \\\hline
$g_{NN\pi}$ &\multicolumn{2}{c}{0.989~\cite{Schutz:1994ue,Ronchen:2012eg}}\\
$g_{\Delta N\pi}$   &\multicolumn{2}{c}{2.12~\cite{Schutz:1994ue,Ronchen:2012eg}}\\
$f_{NN\omega}$  &\multicolumn{2}{c}{0~\cite{Ronchen:2012eg,Machleidt:1987hj}}\\
$\alpha_{BBP}$  &\multicolumn{2}{c}{0.4~\cite{Ronchen:2012eg,Machleidt:1987hj}}\\
$\alpha_{BBV}$  &\multicolumn{2}{c}{1.15~\cite{Ronchen:2012eg,Machleidt:1987hj}}\\
$g_{NN\rho}$    &3.1~\cite{Matsuyama:2006rp}    &3.25~\cite{Schutz:1994ue,Liu:2001ce}\\
$\kappa_{\rho}$  &1.825~\cite{Matsuyama:2006rp}    &6.1~\cite{Schutz:1994ue}\\
$g_{\Delta N\rho}$  &6.08~\cite{Matsuyama:2006rp}    &16.0~\cite{Schutz:1994ue}\\
\hline
\end{tabular*}
\end{table}

In Table~\ref{tab:value-coup}, we present the values for the meson-baryon coupling constants used in our calculations
[see Eqs.~(\ref{eq:coup-relation-begin})-(\ref{eq:coup-relation-end})]. Note that $g_{NN\pi}=0.989$ was determined in
Ref.~\cite{Schutz:1994ue} based on $\pi N$ scattering, where it is found that the $\pi N$ phase shift, scattering length,
and the $\pi N\Sigma $ term were in agreement with the experimental data. We also use $\alpha_{BBP}=0.4$ and $\alpha_{BBV}=1.15$,
taken from an analysis of elastic $N\pi$ scattering~\cite{Ronchen:2012eg}. For the coupling $g_{\Delta N\pi}$ we take the
value 2.12~\cite{Schutz:1994ue} determined from the $\Delta \to N \pi$ decay rate. Besides these well determined parameters,
the value of $\kappa_{\rho}$ can vary in a wider range, e.g., from $\kappa_\rho=1.825$ in Ref.~\cite{Matsuyama:2006rp} to
$\kappa_\rho=6.1$ in Ref.~\cite{Schutz:1994ue}. The values for the $g_{NN\rho}$ coupling cited in these two references do not
vary too much and are also presented in Table~\ref{tab:value-coup}. The difference in values for $\kappa_\rho$ and $g_{NN\rho}$
consequently has an impact on the coupling constant $g_{\Delta N\rho}$~\cite{Matsuyama:2006rp}:
\eq
g_{\Delta N\rho}=\sqrt{\frac{72}{25}}
\frac{g_{NN\rho}(1+\kappa_\rho)}{2m_N}m_\rho\,,
\en
with $g_{\Delta N\rho}=6.08$ in Ref.~\cite{Matsuyama:2006rp} and $g_{\Delta N\rho}=16.0$ in Ref.~\cite{Schutz:1994ue}. Since there
is no way to distinguish the cases of parameter values for $g_{NN\rho}$, $\kappa_\rho$, and $g_{\Delta N\rho}$ we use both sets
in the present calculation. Further details will be discussed in the next section.

Starting from our total effective Lagrangians we generate matrix elements corresponding
to the diagrams of Fig.~\ref{fig:dia-xi-sig/lam}. Their expressions read as follows:
\eq
\mathcal{M}_i = \bar u(p_4) \, \Lambda^{\alpha\beta}_i(p_3,p_4) \, C \bar u^T(p_3) \,
\epsilon_{\alpha\beta}^{(\lambda)}(p)
\quad i=1, 2\,,
\en
where
\eq
\Lambda_1^{\alpha\beta}(p_3,p_4) &=& - g_{d_{N\Omega}} \,
\frac{f_{\Omega\Xi K} \, f_{\Lambda pK}}{m_\pi^2} \,
\int\frac{d^4q}{(2\pi)^4i} \,
q_\mu q_\nu \nonumber\\
&\times& D(q,m_K) \, S^{\nu\alpha}(p_2,m_\Omega) \,
\gamma^\beta \, S(-p_1,m_p) \, \gamma^5 \gamma^\mu
\nonumber\\
&\times& \tilde\Phi\Big(-(p_1-w_{p\Omega}p)^2\Big)
\, \mathcal F(m_t,q)\,, \\
\Lambda_2^{\alpha\beta}(p_3,p_4) &=&  g_{d_{N\Omega}} \,
\frac{g_{\Omega\Xi K^\ast} g_{\Lambda pK^\ast}}{m_\rho} \,
\int\frac{d^4q}{(2\pi)^4i} \,
\Big[ g_{\rho\sigma} q_\tau - g_{\rho\tau} q_\sigma \Big]
\nonumber\\
&\times& \gamma^\tau \gamma^5 \,
S^{\sigma\alpha}(p_2,m_\Omega) \, \gamma^\beta \,
S(-p_1,m_p) \, D^{\mu\rho}(q,m_{K^*}) \nonumber\\
&\times&
\biggl[ \gamma_\mu - i \sigma_{\mu\nu} q^\nu \,
\frac{\kappa_{K^\ast\Lambda p}}{2 m_p}
\biggr] \nonumber\\
&\times&
\tilde\Phi\Big(-(p_1-w_{p\Omega}p)^2\Big) \,
\mathcal F(m_t,q) \,.
\en
where $D(q,m_K) = (q^2-m_K^2)^{-1}$ and $D_{\mu\nu}(q,m_{K^*}) = (-g_{\mu\nu}+q_\mu q_\nu/m_{K^*}^2) \, (q^2-m_{K^*}^2)^{-1}$
are the propagators of the $K$ and $K^*$ mesons, respectively.

A phenomenological dipole form factor
\eq\label{alpha_par}
\mathcal F(m_t,q)=(m_t^2-\Lambda_1^2)^2/(q^2-\Lambda_1^2)^2
\en
is introduced to take into account off-shell effects and the nonlocal structure of the interacting particles~\cite{Cheng:2004ru}.
Here $\Lambda_1=m_t+\alpha\Lambda_{\text{QCD}}$ is a cut-off parameter with $m_t$ being the mass of the exchange particle and the
QCD scale parameter $\Lambda_{\text{QCD}}=0.22$\,GeV. The other two transition amplitudes corresponding to the diagrams in 
Figs.~\ref{fig:dia-xi-sig/lam}(3) and \ref{fig:dia-xi-sig/lam}(4) are generated from the underlying phenomenological Lagrangian
in analogy.

Finally, the total contribution to the matrix element of the $d_{N\Omega}\to\Xi\Lambda$ process is:
\eq
\mathcal{M}_{\text{tot}}(d_{N\Omega}\to\Xi\Lambda)=\mathcal M_1+\mathcal M_2
\en
and for the $d_{N\Omega}\to\Xi\Sigma$ transition
\eq
\mathcal{M}_{\text{tot}}(d_{N\Omega}\to\Xi\Sigma)=\mathcal M_3+\mathcal M_4\,.
\en
The expression for the decay widths of $d_{N\Omega}\to\Xi\Lambda/\Xi\Sigma$ is evaluated as
\eq
\Gamma=\frac{1}{2J+1} \, \frac{|\vec p\,|}
{8 \pi m_{d_{N\Omega}}^2} \, |\overline{{{\mathcal{M}}_{\text{tot}}}}|^2\,,
\en
where $J$ is the total angular momentum of the initial state $d_{N\Omega}$, $\vec p$ is the relative 3-momentum of the final
states in the rest frame of the initial state and the overline denotes the sum of spin polarizations for initial/final states.

\section{Numerical Results \label{section:3}}

\begin{table}[hbt!]
\caption{Masses of the relevant particles (in units of GeV)\cite{Patrignani:2016xqp}.\label{tab:value-mas}}
\vspace{0.2cm}
\begin{tabular*}{\hsize}{@{}@{\extracolsep{\fill}}c|cccccc@{}}
\hline
Particle    &$d_{N\Omega}$       &$p$  &$\Lambda^0$    &$\Sigma^0$ &$\Xi^0$    &$\Omega^-$\\
    \hline
Mass   &2.608    &0.9383 &1.116 &1.193 &1.315  &1.672\\
\hline
\end{tabular*}
\end{table}
First in Table~\ref{tab:value-mas} we summarize the mass values used in the present calculation~\cite{Patrignani:2016xqp}. Although
different masses were predicted for the $d_{N\Omega}$, we choose a mass for the $d_{N\Omega}$ as reliably determined in the recent
lattice calculation, where the binding energy is about 2.46 MeV.

\begin{table}[hbt!]
\caption{Dependence of the coupling $g_{d_{N\Omega}}$ on $\Lambda$.
\label{tab:n-coup-dNOmega}}
 \begin{tabular*}{\hsize}{@{}@{\extracolsep{\fill}}c|cccc@{}}
    \hline
   $\Lambda$\,(GeV)   &0.20    &0.30    &0.40   &0.50 \\\hline
   $g_{d_{N\Omega}}$  &2.38    &2.11   &1.97   &1.88 \\
   \hline
  \end{tabular*}
\end{table}

Values for the coupling $g_{_{d_{N\Omega}}}$ of the $d_{N\Omega}$ bound state to the constituents are generated by the compositeness
condition and are listed in Table III. The values depend on the model parameter $\Lambda$, which is introduced in the correlation
function of Eq.~(\ref{eq:corre-func}) and phenomenologically represents  the distribution of the $N$ and $\Omega$ baryons in the
$d_{N\Omega}$. In Ref.~\cite{Dong:2008mt}, utilizing the same approach for the rather weakly bound deuteron, the parameter $\Lambda$
was deduced to be less than 0.5 GeV. The numerical results for the deuteron electromagnetic form factors of Ref.~\cite{Dong:2008mt}
are in fairly good agreement with data. Based on the similarity between the $d_{N\Omega}$ and the deuteron, we choose four typical
values for the phenomenological cutoff parameter $\Lambda=0.2$, 0.3, 0.4 and 0.5 GeV. The resulting values for
$g_{_{d_{N\Omega}}}$ are 2.38, 2.11, 1.97 and 1.88, respectively (see Table~\ref{tab:n-coup-dNOmega}).

\begin{table}[hbt!]
{\caption{Two-body decay widths of the $d_{N\Omega}$
in keV for different values  of $\Lambda$. The uncertainties of results for a fixed $\Lambda$
reflect the variation in $\alpha$ ranging from 0.9 to 1.1. Coupling constants are taken from Set I. \label{tab:n-toLamXi-2.46}}}
 \begin{tabular*}{\hsize}{@{}@{\extracolsep{\fill}}c|cccc@{}}
    \hline
    Parameters   &\multicolumn{4}{c}{Set I} \\\hline
   $\Lambda$\,(GeV)   &0.20    &0.30    &0.40   &0.50 \\
   \hline
   {$\Lambda^0\Xi^0$ mode} &154$\sim$275 &253$\sim$455  &321$\sim$582   &355$\sim$646\\

   {$\Sigma^0\Xi^0$ mode}  &4.00$\sim$6.61  &6.00$\sim$10.0 &6.75$\sim$11.4 &7.03$\sim$12.0\\
   {$\Sigma^+\Xi^-$ mode}   &8.00$\sim$13.2  &12.0$\sim$20.0 &13.5$\sim$22.8 &14.1$\sim$24.0\\
   \hline
   Total    &166$\sim$295   &271$\sim$485   &341$\sim$616   &376$\sim$682\\
   \hline
  \end{tabular*}
\end{table}

Finally, we have a remaining parameter $\alpha$ in the phenomenological form factor of Eq.~(\ref{alpha_par}). The parameter $\alpha$
cannot be fixed from first principles, instead we choose $\alpha=0.9-1.1$ previously determined from an extended analysis of decay
data on possible baryon-antibaryon bound states (see, e.g., the detailed discussion in Ref.~\cite{Dong:2017rmg}).

In Tables~\ref{tab:n-toLamXi-2.46} and \ref{tab:n-toLamXi-2.46_2}, the partial strong decay widths of the  $d_{N\Omega}\to \Lambda^0\Xi^0$,
$d_{N\Omega}\to \Sigma^0\Xi^0$ and $d_{N\Omega}\to \Sigma^+\Xi^-$ transitions together with their dependence on $\Lambda$ are
displayed. For a fixed value of $\Lambda$ the range in results corresponds to a variation of the parameter $\alpha$ from
0.9 to 1.1 entering in the transition form factor.

Using the values for the coupling constants of Set I we find that the partial strong decay width for
$d_{N\Omega}\to \Lambda^0\Xi^0$ varies from 154$\sim$275 keV to 355$\sim$646 keV, and that for
$d_{N\Omega}\to \Sigma^{0}\Xi^{0}$ from 4.0$\sim$6.61 keV to 7.03$\sim$12.0 keV. Therefore, the mode  $d_{N\Omega}\to \Lambda^0\Xi^0$
dominates over the $d_{N\Omega}\to \Sigma^{0}\Xi^{0}$ decay. From the relations of Eqs.~(\ref{eq:coup-disc-1})
and (\ref{eq:coup-disc-2}) for the couplings, it is clear that $g_{\Lambda p K}$ is much larger than  $g_{\Sigma pK}$,  therefore
resulting in a dominant branching fraction of the $\Lambda\Xi$ mode. In addition, the partial width of the charged $\Sigma^+\Xi^-$ mode
was obtained by isospin symmetry, where isospin breaking effects, like mass differences of charged and neutral baryons, are
not considered. Assuming that the sum of the three partial decay widths results in  the total decay width, we can conclude that
the total decay width of the $d_{N\Omega}$ is in the range of 166$\sim$682 keV. \\

\begin{table}[hbt!]
{\caption{Two-body decay widths of the $d_{N\Omega}$ in keV for different values  of $\Lambda$.  The range of results for $\Lambda$
corresponds to the variation in $\alpha$ from 0.9 to 1.1. Coupling constants are taken from Set II.
\label{tab:n-toLamXi-2.46_2}}
}
 \begin{tabular*}{\hsize}{@{}@{\extracolsep{\fill}}c|cccc@{}}
    \hline
    Parameters   &\multicolumn{4}{c}{Set II} \\\hline
   $\Lambda$\,(GeV)   &0.20    &0.30    &0.40   &0.50 \\
   \hline
   {$\Lambda^0\Xi^0$ mode} &329$\sim$593    &546$\sim$993   &741$\sim$1360   &842$\sim$1550\\
   {$\Sigma^0\Xi^0$ mode}  &10.3$\sim$17.7  &16.1$\sim$27.9 &20.4$\sim$36.0 &22.5$\sim$40.0\\
   {$\Sigma^+\Xi^-$ mode}  &20.6$\sim$35.4  &32.2$\sim$55.8 &40.8$\sim$72.0 &45.0$\sim$80.0\\
   \hline
   Total    &360$\sim$646   &594$\sim$1080  &802$\sim$1470  &910$\sim$1670\\
   \hline
  \end{tabular*}
\end{table}

With the other set of coupling constants (Set II), we found that the partial decay widths for both the $\Lambda\Xi$ and
$\Sigma\Xi$ modes increase by a factor of about two. The obtained partial decay width for the $\Lambda^0\Xi^0$  mode is
from 329 to 1550 keV, and for the $\Sigma^0\Xi^0$  decay width we have values from 10.3 to 40.0 keV when varying $\Lambda$ and
$\alpha$ in the allowed range. For Set II of the coupling constant we conclude that the total decay width of $d_{N\Omega}$ is
in the range of 360$\sim$1670 keV.

\begin{table}[hbt!]
\caption{Decay widths of the processes $d_{N\Omega}\to\Xi^0\Lambda^0$ and
$d_{N\Omega}\to\Xi^0\Sigma^0$ in units of keV.
The binding
energy of the $d_{N\Omega}$ is fixed at 2.46 MeV.
The parameter $\Lambda$ is chosen
to be 0.2 GeV and $\alpha$ is 0.9.
\label{tab:num-individual}}
\vspace{0.2cm}
\begin{tabular*}{\hsize}{@{}@{\extracolsep{\fill}}c|ccc|ccc@{}}    \hline
Parameters  &\multicolumn{3}{c|}{Set I}  &\multicolumn{3}{c}{Set II}\\\hline
Individual contribution   &$K$    &$K^\ast_V$    &$K^\ast_T$    &$K$    &$K^\ast_S$    &$K^\ast_T$\\
   \hline
   {$\Gamma(d_{N\Omega}\to\Xi^0\Lambda^0)$}    &125  &0.713 &0.291  &125    &5.42   &24.8\\
   \hline
   {$\Gamma(d_{N\Omega}\to\Xi^0\Sigma^0)$}    &4.33   &0.146 &0.0444 &4.33   &1.11   &3.78 \\
   \hline
  \end{tabular*}
\end{table}

To check for the contribution of individual diagrams to the processes $d_{N\Omega}\to \Xi\Lambda$ and $d_{N\Omega}\to \Xi\Sigma$
as well as the effect of different coupling values, we analyze the particular results for the partial decay widths as given in
Table~\ref{tab:num-individual}. The detailed results are based on the choice $\Lambda=0.2$\,GeV and $\alpha=0.9$. The entry for
$K$ in Table~\ref{tab:num-individual} represents the contribution from the $K$ meson exchange as shown in Fig.~\ref{fig:tri-decay},
while $K^\ast_V$ and $K^\ast_T$ correspond to the vector and tensor parts of the $K^\ast$ meson exchange contribution.
For the couplings of Set I it is clearly seen that $K$ exchange plays the essential role in both $d_{N\Omega}\to \Xi^0\Lambda^0$
and $d_{N\Omega}\to \Xi^0\Sigma^0$ processes. The contribution of $K^\ast$ exchange is, at least, one order of magnitude smaller,
where the tensor part is much smaller than the vector part. For Set II  the $K$ meson exchange results in the same values for the
partial decay widths since the relevant coupling constants $g_{\Omega\Xi K}$, $f_{\Sigma NK}$ and $f_{\Lambda NK}$ are the same
in the two cases. But now the contribution from $K^\ast$ exchange increases since the coupling constant $g_{\Omega\Xi K^\ast}$
and $\kappa $ are larger, where the tensor contribution dominates over the vector part.

Therefore, we conclude that for both sets of couplings the $K$ meson exchange contribution is dominant for both
$\Lambda\Xi$ and $\Sigma\Xi$ modes; hence, the full decay widths do not change dramatically within the two different sets of
parameter values. The uncertainties in the parameters $\Lambda$ and $\alpha$ obviously have a sizable impact on the calculated
decay widths. The total decay width can reach from a few hundred to above a thousand keV although the transitions of the
$d_{N\Omega}$ to the possible final states occur via a $D$-wave. This analysis of the partial decay widths indicates
that the  process $d_{N\Omega}\to\Lambda\Xi$ dominates in the $d_{N\Omega}$ decays with a branching fraction of around 95\%
independent of the particular parameter choice.

\begin{table}[hbt!]
\caption{Two-body decay width of $d_{N\Omega}\to\Lambda\Xi$ in dependence on
$\Lambda$, while $\alpha$ is set to 1.0.
The mass of $d_{N\Omega}$ is set to $m_{d_{N\Omega}}=2566$\,MeV, which is the same as
in Ref.~\cite{Pang:2004mm} with a corresponding binding
energy of 44.5\,MeV. \label{tab:n-toLamXi-2566}}
\vspace*{0.2cm}
\begin{tabular*}{\hsize}{@{}@{\extracolsep{\fill}}c|cccc@{}}
    \hline
   $\Lambda$\,(GeV)   &0.20    &0.30    &0.40    &0.5\\
   \hline
   Coupling $g_{d_{N\Omega}}$   &13.0   &8.90   &7.16   &6.21\\
   \hline
   \multirow{2}*{$\Gamma(d_{N\Omega}\to\Xi\Lambda)$}  &300  &442 &487    &485\\
   \cline{2-5}
   &Ref.~\cite{Pang:2004mm} &\multicolumn{3}{c}{33.9}\\
   \hline
  \end{tabular*}
\end{table}

The partial decay width for the process $d_{N\Omega}\to\Lambda\Xi$ was also estimated in Ref.~\cite{Pang:2004mm} in the context of a
quark model. They relied on a different mass $M_{{d_{N\Omega}}}=2566$ MeV or on the corresponding binding energy of $E_b=44.5$ MeV,
and used quark rearrangement for the decay mechanism. Quantitatively the results for quark rearrangement and for the  meson exchange
process are very different. To directly compare our results with theirs, we indicate our results for the partial decays width of
$d_{N\Omega}\to\Lambda\Xi$ for the mass of the $d_{N\Omega}$ being 2566 MeV as well. Results are listed in
Table~\ref{tab:n-toLamXi-2566} for the comparison. Here we take the averaged value $\alpha=1.0$ and the coupling parameters of Set I.
For parameter Set II our results for the decay width are about twice as large compared to those of Set I.

Compared to the numbers obtained for the mass of the lattice prediction the values of the coupling constant $g_{d_{N\Omega}}$
increase. This is a natural result  since the larger binding also corresponds to a stronger interaction reflected by
$g_{d_{N\Omega}}$. Our results are almost one order of magnitude larger than the one of the quark model~\cite{Pang:2004mm}
although the input mass is the same. It should be addressed that another estimate for the total decay width of the ${d_{N\Omega}}$
was also obtained in the meson exchange model~\cite{Sekihara:2018tsb}. With a binding energy of 0.1 MeV the decay width of
$\Lambda\Xi$ and $\Sigma\Xi$ is 1.5 MeV.

Here, we want to emphasize that in the present work the $d_{N\Omega}$ was assigned as a pure $N\Omega$ molecular state, while such a
pure bound state was supported by the lattice calculation\cite{Iritani:2018sra}. The current results are the decay properties of
such an $N\Omega$ molecular state. On the other hand, one can not exclude other components, e.g., $\Lambda\Xi^\ast$ in the $d_{N\Omega}$.
And the additional components may have an effect on the decays of the $d_{N\Omega}$. This issue will be studied elsewhere.\\

From our results in Table III we find that the $\Lambda^0\Xi^0$ decay mode completely dominates the total decay width. This phenomenon
occurs because of the large coupling  constant $g_{\Lambda pK}$.  The final state $\Lambda^0\Xi^0$ is not easily observed in experiment
since the final hadrons dominantly decay through weak decay processes. For example, the $\Lambda^0$ can decay to $p\pi^-$ and
$n\pi^0$, where the $p\pi^-$ mode is preferred to reconstruct the $\Lambda^0$. Moreover, $\Xi^0$ can be reconstructed by
the three-body final state of $p\pi^-\pi^0$ because of the decay chain $\Xi^0\to\Lambda^0\pi^0\to p\pi^-\pi^0$.

The $d_{N\Omega}$ can also decay to $N\Lambda K$ via the weak $\Omega\Lambda K$ vertex, where the final state particles can be
easily observed. However, this weak decay proportional to $G_F$ is strongly suppressed. The dominant strong decay process
$d_{N\Omega}\to\Lambda\Xi$ is clearly the signal of a possible $d_{N\Omega}$ to be searched for. If the accumulated experimental
data sample for hyperons is large enough, one may see the signal of the $d_{N\Omega}$ in the $\Lambda\Xi$ invariant mass spectrum.
We also know that the RHIC experiment has already presented a positive result for $d_{N\Omega}$ via indirect measurements.
Future experiments are expected to provide more direct and precise evidence for the existence of the $d_{N\Omega}$.

\section{Summary}
\label{section:4}

The $d_{N\Omega}$ stands for a bound, minimal, six-quark configuration with baryon number $B=2$ and strangeness $S=-3$. Because of its
weak binding it is analogous to the deuteron, which is an experimentally confirmed baryon-baryon bound state. The $d_{N\Omega}$
was predicted in many theoretical works and, in particular, by lattice calculations. There are also some hints for the existence
of the $d_{N\Omega}$ from recent experimental approaches. In the present work, we give an analysis of the strong decays  of the
$d_{N\Omega}$ based on the use of phenomenological Lagrangian approach,  in which the $d_{N\Omega}$ is assumed to be a loosely bound
state. Here, we simply use the lattice prediction for the binding energy.

All possible strong two-body decay modes of the $d_{N\Omega}$ are calculated. In the calculation two sets of coupling parameters
are employed. We find that the total decay width of the $d_{N\Omega}$ is in the range of a few hundred keV up to just above 1 MeV
although the transitions to the two modes proceed through the $D$-wave. Independent of the particular choice of parameters
the $d_{N\Omega}\to\Lambda\Xi$ process dominates completely and almost captures the total branching fraction. A search for
the $d_{N\Omega}$ in the $\Lambda\Xi$ invariant mass spectrum can provide direct evidence for its existence. Finally, we would like
to point out that more theoretical efforts are needed to understand the structure of the $d_{N\Omega}$ exotic state as well as to
search for other possible candidates in the baryon-baryon molecular state family.

\section*{Acknowledgement}

This work is supported, in part, by the National Natural Science Foundation of China (NSFC)
under Grant Nos.~11947224, 11975245 and 11775050, by the fund provided to the Sino-German CRC 110
"Symmetries and the Emergence of Structure in QCD" project by the NSFC under Grant No.~11621131001,
by the Key Research Program of Frontier Sciences, CAS, Grant No. Y7292610K1,
by the Fundamental Research Funds for the Central Universities, and by the China Postdoctoral Science
Foundation under Grant No.~2019M650843.
This work was funded by the Carl Zeiss Foundation under Project ``Kepler Center f\"ur Astro- und
Teilchenphysik: Hochsensitive Nachweistechnik zur Erforschung des unsichtbaren Universums
(Gz: 0653-2.8/581/2)'', by ``Verbundprojekt 05A2017 - ¡ªCryogenic Rare
Event Search with Superconducting Thermometers(CRESST)-XENON: Direkte Suche
nach Dunkler Materie mit XENON1T/nT und CRESST-III. Teilprojekt 1
(F\"orderkennzeichen 05A17VTA)'', by "Verbundprojekt 05P2018 - Ausbau von ALICE am LHC: Jets und partonische Struktur von Kernen"(F\"{o}rderkennzeichen: 05P18VTCA1), Agencia Nacional de Investigaci\'{¨®}n y Desarrollo (ANID) PIA/APOYO AFB180002 (Chile) and
by FONDECYT (Chile) under Grant No. 1191103. The authors thank the unknown Referee for the valuable comments. YBD thanks Institute of Theoretical Physics, T\"ubingen University
for the hospitality and the
support of Alexander von Humboldt foundation, and Jialun Ping for useful discussions.

\end{document}